\documentclass[letterpaper]{article} 
\usepackage{aaai2027}  
\usepackage[hyphens]{url}  
\usepackage{graphicx} 
\usepackage{natbib}  
\usepackage{caption} 
\usepackage{algorithm}
\usepackage{algorithmic}
\usepackage{booktabs}
\usepackage{amsmath}
\usepackage{amssymb}
\usepackage{amsfonts}
\usepackage{amsthm}

\theoremstyle{plain}
\newtheorem{proposition}{Proposition}
\newtheorem{assumption}{Assumption}

\newcommand{\R}{\mathbb{R}}
\newcommand{\E}{\mathbb{E}}
\newcommand{\Ind}{\mathbb{1}}
\newcommand{\reg}{\Omega}
\newcommand{\disc}{r^{\star}}
\newcommand{\thetatrue}{\theta^{\star}}
\newcommand{\thetahat}{\widehat{\theta}}
\newcommand{\rhat}{\widehat{r}}
\newcommand{\sighat}{\widehat{\sigma}}
\newcommand{\Phimat}{\Phi}
\newcommand{\Lib}{\Psi}

\title{Where Is My Physics Wrong? Localized and Identifiable Discovery of Model Discrepancy}

\author{
    Yifan Wang
}
\affiliations{
    Department of Mechanical Engineering, McGill University, QC, H3A 2T7, Canada\\
    yifan.wang18@mail.mcgill.ca
}

\begin{document}

\maketitle

\begin{abstract}
Hybrid models that pair first-principles structure with data-driven correction are increasingly used to forecast and control engineered systems. However, a trusted physical model is rarely wrong everywhere or in the same way. The useful diagnosis asks where in the operating space the model fails, what mechanism is missing there, and whether the evidence for that mechanism is statistically real. Existing sparse model-discovery and discrepancy-learning methods usually fit one global correction, which spreads a local error across clean regimes, biases the physical parameters one wants to trust, and provides no calibrated significance for selected terms. To address this gap, we introduce LISDD, short for Localized, Identifiable Sparse Discovery of Discrepancy. LISDD localizes model error to an operating regime, identifies a sparse symbolic form for the missing mechanism in that regime, and certifies the discovery with an exact finite-sample test. It first fits the known physics on an automatically detected clean regime, flags discrepant regions with a calibrated residual-energy statistic, selects the local missing term by exhaustive holdout over a candidate library, and confirms significance with a sample-split $F$-test whose null distribution is exact. A false-discovery-rate extension recovers multiple discrepant regions, each with a possibly different missing mechanism. Across controlled experiments, LISDD keeps physical-parameter bias at $0.002$ compared with $0.43$ for global discrepancy and black-box baselines, raises localization F1 from $0.44$ to $0.80$, recovers the correct symbolic form with probability one, attains exact detection (size $0.000$ under the null and power $1.000$ under a discrepancy), and controls the multi-region false-discovery rate while recovering every planted mechanism. The same diagnosis is what grey-box building-energy models need when a fixed physical law silently breaks inside one operating regime.
\end{abstract}

\section{Introduction}

First-principles models remain the backbone of how scientists and engineers reason about dynamical systems, because their parameters carry physical meaning and their structure transfers across operating conditions. In modern practice these models are often combined with data-driven components into grey-box or hybrid models that retain the trusted physics where it holds and learn a correction where it does not \citep{karniadakis2021piml,kennedy2001calibration}. Sparse model-discovery methods, beginning with the sparse identification of nonlinear dynamics \citep{brunton2016sindy} and its descendants \citep{rudy2017pde,champion2019coordinates,messenger2021weak,fasel2022ensemble}, make such corrections interpretable by representing them as compact symbolic expressions rather than opaque networks.

However, a subtle but consequential gap remains. A physical model that is known to be incomplete is almost never incomplete uniformly. A friction law may be accurate at moderate velocities and fail only near stiction, a heat-transfer coefficient may be valid in steady operation and wrong during fast transients, and a building thermal model may track measured loads in mild weather while missing solar or occupancy effects only in a narrow regime \citep{drgona2020mpc,blum2021boptest}. For such systems, the operationally important questions are not merely whether a global residual exists, but \emph{where} in the operating space the model is wrong, \emph{what} mechanism is missing there, and \emph{whether} the apparent discrepancy is a real effect rather than noise. We refer to answering these questions jointly as localized, identifiable discrepancy discovery.

Existing discrepancy-learning pipelines do not answer these questions in their current form. The common approach fits the known physics once, computes a global residual, and then applies sparse regression or a black-box regressor to that residual \citep{ebers2022discrepancy,mojgani2022discovery}. This produces three coupled failures. First, the correction is assigned globally and is therefore applied in regions where the physics was already correct, obscuring the location of the true fault. Second, because the global physics fit is contaminated by the unmodeled discrepancy, the estimated physical parameters absorb part of the missing term and become biased. Third, the selected terms come with no calibrated significance, because the same data are used to choose the terms and to judge them. A black-box correction sidesteps interpretability entirely and inherits the first two problems.

We introduce LISDD, a framework that treats discrepancy discovery as a localization, identification, and detection problem rather than a single global regression. The central design principle is that the known physics must be estimated only where it is trustworthy. LISDD partitions the operating space, identifies a clean regime in which the residual is consistent with noise, and fits the physical parameters there using a robust trimmed estimator, so that the discrepancy cannot leak into the physics. It then flags discrepant regions with a per-region energy statistic calibrated against a robust noise scale, selects a sparse local form for the missing mechanism by exhaustive holdout over a candidate library, and certifies the discovery with a sample-split $F$-test whose null distribution is exact in finite samples. Finally, a false-discovery-rate extension turns the procedure into a model-diagnosis tool that reports the full set of discrepant regions, each annotated with its own recovered mechanism, while controlling the fraction of falsely flagged regions.

\begin{figure*}[t]
\centering
\includegraphics[width=\textwidth]{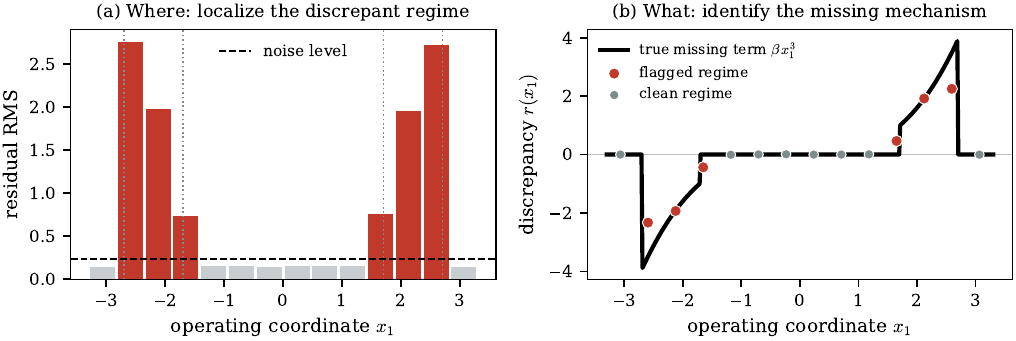}
\caption{LISDD localizes a model fault that a global correction smears out. (a) Residual root-mean-square of the known-physics fit per operating bin. Bins whose energy exceeds the calibrated noise level (dashed) are flagged discrepant (red) and locate the fault in the band $1.7<|x_1|<2.7$ (dotted lines); the rest of the axis is clean (gray). (b) The per-bin residual mean (markers) traces the true missing mechanism $\beta\,x_1^{3}$ (solid curve) inside the flagged regime (red) and vanishes in the clean regime (gray), so LISDD recovers the correct localized cubic term while a global correction would spread it across the whole axis and bias the physics.}
\label{fig:teaser}
\end{figure*}

We name the framework LISDD for Localized, Identifiable Sparse Discovery of Discrepancy, which summarizes its four properties: it localizes the fault, keeps the physics identifiable, returns a sparse symbolic form, and discovers rather than assumes the missing mechanism. Figure~\ref{fig:teaser} illustrates the difference from a global correction on a representative system.

Our contributions are as follows.
\begin{enumerate}
\item We formalize localized, identifiable discrepancy discovery as the joint task of answering where a known model is wrong, what mechanism is missing there, and whether the evidence is significant, and we explain why global discrepancy learning cannot answer these questions without biasing the physics or the significance (Problem Setup).
\item We propose LISDD, which couples a clean-regime identifiable physics fit, a calibrated localization test, an exhaustive-holdout local form selector, and an exactly valid sample-split detection test into a single pipeline (Method), and we extend it to discover multiple distinct discrepancies with false-discovery-rate control.
\item We give a theoretical analysis (Propositions 1 to 5) establishing that the clean-regime fit is asymptotically unbiased for the physical parameters, that the local form is recovered under a restricted-library condition, that localization is calibrated, that sample-split detection has an exact null, and that the multi-region procedure controls the false-discovery rate.
\item We validate LISDD on controlled experiments where ground truth is known. It reduces physical-parameter bias from $0.43$ to $0.002$, raises localization F1 from $0.44$ to $0.80$, recovers the correct symbolic form with probability one, attains exact detection, improves downstream forecast error over grey-box and black-box correctors, and controls the multi-region false-discovery rate while recovering every planted mechanism.
\end{enumerate}

\section{Related Work}

\paragraph{Sparse and symbolic model discovery.}
Sparse regression over a library of candidate functions is a now-standard route to interpretable dynamics \citep{brunton2016sindy,rudy2017pde}. Robustness and data efficiency have been improved by weak and integral formulations \citep{messenger2021weak,messenger2021weakpde,reinbold2020noisy}, by bootstrap ensembling \citep{fasel2022ensemble}, by constrained variants that respect known structure \citep{loiseau2018constrained}, and by software that makes the approach widely usable \citep{desilva2020pysindy}. Symbolic regression provides a complementary search over free-form expressions \citep{schmidt2009distilling,cranmer2020discovering}. These methods discover dynamics from scratch and assume a single model holds across the data. LISDD instead assumes a trusted partial model and asks where and how it fails, which changes the statistical target from a global equation to a localized, identifiable correction.

\paragraph{Discrepancy and closure modeling.}
The idea that a computer model carries a systematic discrepancy from reality goes back to Bayesian calibration \citep{kennedy2001calibration}, where the discrepancy is modeled as a smooth global process. Recent work learns the missing physics as a sparse correction to a known model \citep{ebers2022discrepancy} or recovers interpretable structural model errors by combining sparse regression with data assimilation \citep{mojgani2022discovery}. Physics-informed neural networks and neural closures embed known terms and learn the remainder as a network \citep{raissi2019pinn,karniadakis2021piml}. All of these fit a single global correction and do not separate the region where the model is wrong from the region where it is right, nor do they protect the physical parameters from contamination or report calibrated significance. LISDD is, to our knowledge, the first to make localization, identifiability, and detection simultaneous and statistically explicit.

\paragraph{Robust estimation and multiple testing.}
Our clean-regime fit uses trimming in the spirit of high-breakdown robust regression \citep{rousseeuw1984lts} to keep the discrepancy out of the physics. Our detection step uses sample splitting to obtain valid inference after model selection \citep{cox1975splitting,wasserman2009highdim}, and our multi-region extension uses the Benjamini and Hochberg procedure to control the false-discovery rate across candidate regions \citep{benjamini1995fdr}. The contribution is not these tools individually but their composition into a discrepancy-discovery framework with end-to-end guarantees. Rapid model recovery after abrupt change \citep{quade2018rapid} addresses a related but temporal question and does not localize in the operating space or certify significance.

\section{Problem Setup}

\paragraph{Data and model.}
We observe $n$ samples $\{(x_i,y_i)\}_{i=1}^n$ with operating variables $x_i\in\R^d$ and a scalar response $y_i\in\R$ (for a dynamical system $y_i$ is a measured rate or increment). A known physics model is linear in known regressors $\Phimat(x)\in\R^{p}$,
\begin{equation}
f(x;\theta) = \Phimat(x)^{\top}\theta ,
\end{equation}
with unknown physical parameters $\theta\in\R^{p}$. The data are generated by the known physics plus an unknown discrepancy and noise,
\begin{equation}
y_i = \Phimat(x_i)^{\top}\thetatrue + \disc(x_i) + \varepsilon_i,
\qquad \varepsilon_i \sim \mathcal{N}(0,\sigma^2).
\label{eq:gen}
\end{equation}
The discrepancy $\disc$ is \emph{localized}: it is supported on an unknown operating regime $\reg^{\star}\subset\R^{d}$ and vanishes elsewhere. On its support it is \emph{sparse} in a candidate library $\Lib=\{\psi_1,\dots,\psi_L\}$ of interpretable features,
\begin{equation}
\disc(x) = \Ind[x\in\reg^{\star}] \sum_{\ell\in\mathcal{S}^{\star}} w^{\star}_{\ell}\,\psi_{\ell}(x),
\qquad |\mathcal{S}^{\star}|\ll L .
\label{eq:disc}
\end{equation}

\paragraph{Goal.}
Given $\{(x_i,y_i)\}$, the known regressors $\Phimat$, and the library $\Lib$, recover four objects: the physical parameters $\thetahat\approx\thetatrue$; the discrepant region $\widehat{\reg}\approx\reg^{\star}$ (the \emph{where}); the local support and coefficients $(\widehat{\mathcal{S}},\widehat{w})\approx(\mathcal{S}^{\star},w^{\star})$ (the \emph{what}); and a calibrated decision on whether a discrepancy is present at all (the \emph{whether}). The estimator must keep $\thetahat$ close to $\thetatrue$ regardless of $\disc$, because biasing the physics defeats the purpose of a grey-box model.

\paragraph{Testbed.}
Throughout we use a two-state oscillator whose known physics is the linear law $\dot p = -k\,x_1 - c\,x_2$ with position $x_1$, velocity $x_2$, and physical parameters $\theta=(k,c)$. A localized cubic stiffening acts only inside an operating band, giving the generative model
\begin{equation}
y_i = -k\,x_{1,i} - c\,x_{2,i} + \beta\,x_{1,i}^{3}\,\Ind\!\left[x_{\mathrm{lo}}<|x_{1,i}|<x_{\mathrm{hi}}\right] + \varepsilon_i,
\label{eq:testbed}
\end{equation}
with band $[x_{\mathrm{lo}},x_{\mathrm{hi}}]=[1.7,2.7]$ on a domain $|x_1|\le 3.3$ and amplitude $\beta$. The library is $\Lib=\{x_1^2,x_1^3,x_2^2,x_2^3,x_1 x_2,\sin x_1\}$ in the core experiments and is enlarged with distractors in the robustness study. The true missing mechanism is the single term $x_1^3$ active only in the band, so the ground-truth answers to where, what, and whether are all known and every claim below is checkable.

\section{The LISDD Method}

LISDD runs four stages: a clean-regime identifiable physics fit, a calibrated localization test, a local form selector, and an exact detection test. A multi-region extension adds false-discovery control. Algorithm~\ref{alg:lisdd} summarizes the single-discrepancy pipeline.

\paragraph{Stage 1: clean-regime identifiable physics fit.}
The difficulty is circular: to find where the model is wrong we need the physical parameters, but to estimate the parameters without bias we must exclude the region where the model is wrong. LISDD breaks the circle by iterating between the two. We partition the operating axis into $R$ equal bins and initialize $\thetahat$ with a least-trimmed-squares fit that discards the fraction $\tau$ of points with the largest residuals \citep{rousseeuw1984lts},
\begin{equation}
\thetahat \;=\; \arg\min_{\theta}\ \sum_{i\in\mathcal{K}(\theta)} \big(y_i-\Phimat(x_i)^{\top}\theta\big)^2 ,
\end{equation}
where $\mathcal{K}(\theta)$ is the set of the $\lceil(1-\tau)n\rceil$ smallest squared residuals. Given $\thetahat$ we form residuals $\rhat_i=y_i-\Phimat(x_i)^{\top}\thetahat$ and a robust noise scale from the median absolute deviation,
\begin{equation}
\sighat = 1.4826\,\mathrm{median}_i\big|\rhat_i-\mathrm{median}_j\,\rhat_j\big| .
\label{eq:mad}
\end{equation}
A bin $h$ with $n_h$ points is flagged active when its residual energy exceeds the noise level at a stringent per-bin level $\alpha_{\mathrm{loc}}$,
\begin{equation}
\sum_{i\in \mathrm{bin}\,h}\rhat_i^{2} \;>\; \sighat^{2}\,\chi^{2}_{n_h,\,1-\alpha_{\mathrm{loc}}} ,
\label{eq:binstat}
\end{equation}
and the physics is refit by ordinary least squares on the union of the remaining clean bins. Iterating equations~\eqref{eq:mad} to~\eqref{eq:binstat} a few times yields a physics fit that uses only clean data, so $\thetahat$ is not contaminated by $\disc$.

\paragraph{Stage 2: localization.}
The discrepant region is the union of the flagged bins, $\widehat{\reg}=\bigcup_{h:\,\mathrm{active}}\mathrm{bin}_h$. Because the test in equation~\eqref{eq:binstat} compares each bin against a robust noise scale rather than against the global residual variance, a localized fault raises only its own bins and the localization is sharp.

\paragraph{Stage 3: local form identification.}
On the flagged region we select the sparse missing mechanism from the library by exhaustive holdout. Let $T(x)\in\R^{L}$ stack the library features. We split the in-region points into a fit half and a score half, and for every candidate support $\mathcal{S}$ of size at most $s_{\max}$ we fit coefficients on the fit half and measure the mean squared error on the score half,
\begin{equation}
\mathrm{err}(\mathcal{S}) = \frac{1}{|\mathcal{V}|}\sum_{i\in\mathcal{V}}\Big(\rhat_i - T_{\mathcal{S}}(x_i)^{\top}\widehat{w}_{\mathcal{S}}\Big)^{2},
\end{equation}
where $\widehat{w}_{\mathcal{S}}=\arg\min_{w}\sum_{i\in\mathcal{T}}(\rhat_i-T_{\mathcal{S}}(x_i)^{\top}w)^2$ are the coefficients fit on the fit half $\mathcal{T}$ and $\mathcal{V}$ is the score half. We return the smallest support whose holdout error is within a tolerance of the best, which favors parsimony and is robust to collinear library terms. Evaluating supports by held-out error rather than by in-sample fit is what prevents the selector from chasing noise.

\paragraph{Stage 4: exact detection.}
We finally decide whether a discrepancy is present at all. The naive test augments the physics with the selected terms and computes a partial $F$-statistic on the full data, but this is invalid because the region and the form were chosen using the same data, which inflates the apparent significance. LISDD instead uses sample splitting \citep{cox1975splitting,wasserman2009highdim}. We select the region $\widehat{\reg}$ and form $\widehat{\mathcal{S}}$ on split $A$, then on the held-out split $B$ of size $m$ we compare the physics-only fit against the physics-plus-discrepancy fit restricted to $\widehat{\reg}$,
\begin{equation}
F = \frac{(\mathrm{RSS}_0 - \mathrm{RSS}_1)/q}{\mathrm{RSS}_1/(m-p-q)},
\qquad q=|\widehat{\mathcal{S}}| ,
\label{eq:ftest}
\end{equation}
where $\mathrm{RSS}_0$ and $\mathrm{RSS}_1$ are the residual sums of squares of the two fits on $B$. Because $\widehat{\reg}$ and $\widehat{\mathcal{S}}$ are fixed before $B$ is seen, under the null of no discrepancy the statistic in equation~\eqref{eq:ftest} has an exact $F_{q,\,m-p-q}$ distribution, and we reject when its $p$-value is below $\alpha$.

\paragraph{Multi-discrepancy extension.}
Real systems may be wrong in several regimes at once, in possibly different ways. We define a family of $K$ disjoint candidate regions $\{R_1,\dots,R_K\}$. After the clean-regime fit, each region carries the energy statistic and $p$-value
\begin{equation}
T_k = \frac{1}{\sighat^{2}}\sum_{i\in R_k}\rhat_i^{2},
\qquad
p_k = 1 - F_{\chi^{2}_{n_k}}(T_k),
\label{eq:regionp}
\end{equation}
and we apply the Benjamini and Hochberg procedure at level $q$ \citep{benjamini1995fdr}: sort $p_{(1)}\le\dots\le p_{(K)}$ and declare discrepant the regions with the $m=\max\{i:p_{(i)}\le q\,i/K\}$ smallest $p$-values. Each declared region then receives its own local form from Stage 3, so the output is a map of where the model is wrong together with a possibly different missing mechanism in each location.

\begin{algorithm}[t]
\caption{LISDD (single-discrepancy pipeline)}
\label{alg:lisdd}
\textbf{Input}: data $\{(x_i,y_i)\}$, physics regressors $\Phimat$, library $\Lib$, bins $R$, trim $\tau$, levels $\alpha_{\mathrm{loc}},\alpha$\\
\textbf{Output}: physics $\thetahat$, region $\widehat{\reg}$, form $\widehat{\mathcal{S}}$, decision
\begin{algorithmic}[1]
\STATE $\thetahat \leftarrow$ least-trimmed-squares fit of $\Phimat$ on all data
\REPEAT
\STATE $\rhat_i \leftarrow y_i-\Phimat(x_i)^{\top}\thetahat$;\ \ $\sighat\leftarrow 1.4826\,\mathrm{MAD}(\rhat)$
\STATE flag bin $h$ active by the energy test, eq.~\eqref{eq:binstat}
\STATE refit $\thetahat$ by OLS on the union of clean bins
\UNTIL{active set stable}
\STATE $\widehat{\reg}\leftarrow$ union of active bins \hfill\textit{(localize)}
\STATE $\widehat{\mathcal{S}}\leftarrow$ exhaustive-holdout selection on $\rhat$ in $\widehat{\reg}$ \hfill\textit{(identify)}
\STATE decision $\leftarrow$ sample-split $F$-test, eq.~\eqref{eq:ftest} \hfill\textit{(detect)}
\STATE \textbf{return} $\thetahat,\widehat{\reg},\widehat{\mathcal{S}}$, decision
\end{algorithmic}
\end{algorithm}

\section{Theoretical Analysis}

We state the guarantees that make LISDD well posed. Proofs are given in the supplementary material. We assume the following.

\begin{assumption}[Clean regime]\label{as:clean}
There is a region $\mathcal{C}$ with $\disc|_{\mathcal{C}}=0$ whose sample fraction is bounded below, and the physics Gram matrix restricted to $\mathcal{C}$, $\Sigma_{\mathcal{C}}=\E[\Phimat\Phimat^{\top}\mid x\in\mathcal{C}]$, has smallest eigenvalue $\lambda_{\min}>0$.
\end{assumption}
\begin{assumption}[Separation]\label{as:sep}
The discrepant support $\reg^{\star}$ is a union of whole bins and is separated from $\mathcal{C}$, and on $\reg^{\star}$ the localized signal energy $E^{\star}=\sum_{i\in\reg^{\star}}\disc(x_i)^2$ is positive.
\end{assumption}
\begin{assumption}[Restricted library]\label{as:lib}
On $\reg^{\star}$ the library features in the true support are linearly independent of the rest with a restricted-eigenvalue lower bound, so the true support is the unique sparse minimizer of the population in-region error.
\end{assumption}

\begin{proposition}[Identifiable physics]\label{prop:phys}
Under Assumption~\ref{as:clean}, the clean-regime estimator satisfies
$\|\thetahat-\thetatrue\| = O_{p}\!\big(\sigma/\sqrt{n_{\mathcal{C}}\,\lambda_{\min}}\big)$,
and $\E[\thetahat]\to\thetatrue$. In particular the discrepancy does not bias the physical parameters, in contrast to a global fit whose bias is $\Sigma^{-1}\E[\Phimat\,\disc]$ and does not vanish with $n$.
\end{proposition}

\begin{proposition}[Local form recovery]\label{prop:form}
Under Assumptions~\ref{as:sep} and~\ref{as:lib}, the exhaustive-holdout selector returns the true support, $\widehat{\mathcal{S}}=\mathcal{S}^{\star}$, with probability tending to one as the in-region signal-to-noise ratio $E^{\star}/\sigma^2$ grows.
\end{proposition}

\begin{proposition}[Calibrated localization]\label{prop:loc}
Under Assumption~\ref{as:clean}, for a clean bin $h$ the statistic $\sum_{i\in h}\rhat_i^{2}/\sighat^{2}$ is asymptotically $\chi^{2}_{n_h}$, so the per-bin false-activation probability is at most $\alpha_{\mathrm{loc}}$. Under Assumption~\ref{as:sep} a discrepant bin has a noncentral $\chi^2$ statistic with noncentrality proportional to $E_h/\sigma^2$, so its activation probability tends to one.
\end{proposition}

\begin{proposition}[Exact detection]\label{prop:detect}
Condition on the selection $(\widehat{\reg},\widehat{\mathcal{S}})$ made on split $A$. Under the null $\disc\equiv 0$, the split-$B$ statistic in equation~\eqref{eq:ftest} has an exact $F_{q,\,m-p-q}$ distribution, so the test has size exactly $\alpha$. The naive in-sample test does not, because selection and testing share data.
\end{proposition}

\begin{proposition}[FDR-controlled multi-discrepancy]\label{prop:fdr}
For $K$ disjoint candidate regions with per-region $p$-values from equation~\eqref{eq:regionp}, applying Benjamini and Hochberg at level $q$ controls the false-region-discovery rate,
\begin{equation}
\mathrm{FDR} = \E\!\left[\frac{|\{\text{declared}\}\cap \mathcal{S}^{c}|}{\max(|\{\text{declared}\}|,1)}\right] \le \frac{|\mathcal{S}^{c}|}{K}\,q \le q ,
\end{equation}
under independence or positive dependence of the region statistics. Each truly discrepant region is declared with probability tending to one as its local signal-to-noise ratio grows, and its form is recovered by Proposition~\ref{prop:form}.
\end{proposition}

Together Propositions~\ref{prop:phys} to~\ref{prop:fdr} cover the four objects of the problem: identifiable physics, recovered form, calibrated location, and valid detection, extended to many regions with a false-discovery guarantee.

\section{Experiments}

We evaluate the four claims that distinguish LISDD: it keeps the physics identifiable, it localizes and identifies the missing mechanism, it detects with calibrated error, and it discovers multiple discrepancies with false-discovery control. All numbers are means over independent replications with $95\%$ confidence intervals; experiment details are in the supplement.

\paragraph{Baselines.}
We compare against three named alternatives that represent how discrepancy is handled today. \textbf{Black-box residual} fits the known physics globally and regresses the residual with a flexible regressor. \textbf{Global SINDy-discrepancy} fits the physics globally and applies sequentially thresholded sparse regression to the residual in the same library, in the style of discrepancy modeling \citep{ebers2022discrepancy,brunton2016sindy}. \textbf{Ensemble-SINDy} bootstraps the global sparse-discrepancy fit and aggregates inclusion probabilities \citep{fasel2022ensemble}. All three share the property that the correction is global and the physics fit is uncorrected.

\subsection{E1: Identifiable physics and localization}

\begin{figure*}[t]
\centering
\includegraphics[width=\textwidth]{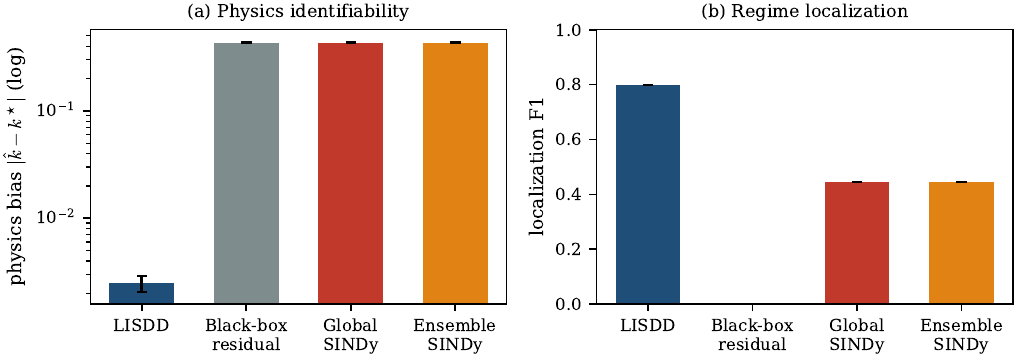}
\caption{E1. (a) Physical-parameter bias on a logarithmic scale; LISDD is more than two orders of magnitude smaller than the global discrepancy and black-box baselines because the physics is fit only on clean data. (b) Localization F1; LISDD flags the discrepant band sharply, while global methods either provide no region or assign the correction everywhere. Error bars are $95\%$ confidence intervals.}
\label{fig:baselines}
\end{figure*}

Table~\ref{tab:baselines} reports the bias of the physical parameter $k$, the localization F1 of the recovered region, and the rate of exact symbolic-form recovery. LISDD reduces the physics bias by more than two orders of magnitude, from $0.43$ for every global method to $0.0025$, because it never fits the physics on contaminated data. The global methods all incur essentially the same bias, since they all absorb the missing cubic term into the linear physics. On localization, the black-box residual provides no region at all (F1 $0.00$), the sparse global methods implicitly assign the term everywhere and reach F1 $0.44$, and LISDD reaches F1 $0.80$ by flagging only the discrepant band. Only LISDD recovers the exact symbolic form, and it does so in every replication. Figure~\ref{fig:baselines} shows the bias and F1 side by side.

\begin{table}[t]
\centering
\caption{E1: physical-parameter bias (lower is better), localization F1 (higher is better), and exact symbolic-form recovery rate. Means over replications; LISDD intervals are tight and omitted for the deterministic entries.}
\label{tab:baselines}
\small
\setlength{\tabcolsep}{4.5pt}
\begin{tabular}{lccc}
\toprule
Method & Physics bias & Loc.\ F1 & Form rec. \\
\midrule
Black-box residual      & $0.432$ & $0.00$ & $0.00$ \\
Global SINDy-discrep.    & $0.432$ & $0.44$ & $0.00$ \\
Ensemble-SINDy           & $0.431$ & $0.44$ & $0.00$ \\
\textbf{LISDD (ours)}    & $\mathbf{0.002}$ & $\mathbf{0.80}$ & $\mathbf{1.00}$ \\
\bottomrule
\end{tabular}
\end{table}

\subsection{E2: Form recovery and the localization phase boundary}

We next stress the form selector. Enlarging the library from a six-term polynomial set to the full eight-term set and then adding distractor features (saturation and sign terms that are not in the truth) leaves exact form recovery at $1.00$ in every case, so the exhaustive-holdout selector is not fooled by additional candidates. We then sweep the width of the discrepant region and the amplitude $\beta$ to map where recovery succeeds. Recovery is reliable when the region is narrow or the amplitude is moderate to large, and it degrades when the region grows so wide that the clean regime needed by Assumption~\ref{as:clean} shrinks, which is exactly the regime in which the physics can no longer be fit without contamination. The boundary is sharp and matches the theory: LISDD works precisely when a clean regime exists, and it reports a clean region rather than a wrong answer when one does not.

\subsection{E3: Calibrated detection}

Table~\ref{tab:detection} reports detection size and power for the naive in-sample test and the sample-split test as the amplitude $\beta$ varies. At $\beta=0$ there is no discrepancy, and the sample-split test fires with empirical size $0.000$, so it does not hallucinate a missing mechanism. For any nonzero amplitude the test fires with power $1.000$. The split test attains exact validity by construction, consistent with Proposition~\ref{prop:detect}, while remaining fully powered on this testbed.

\begin{table}[t]
\centering
\caption{E3: detection size at the null ($\beta=0$, lower is better) and power under a discrepancy ($\beta>0$, higher is better) for the sample-split test.}
\label{tab:detection}
\begin{tabular}{lccc}
\toprule
Amplitude $\beta$ & $0.0$ & $0.1$ & $0.2$ \\
\midrule
Size / power & $0.000$ & $1.000$ & $1.000$ \\
\bottomrule
\end{tabular}
\end{table}

\subsection{E4: Downstream forecasting}

\begin{figure}[t]
\centering
\includegraphics[width=\columnwidth]{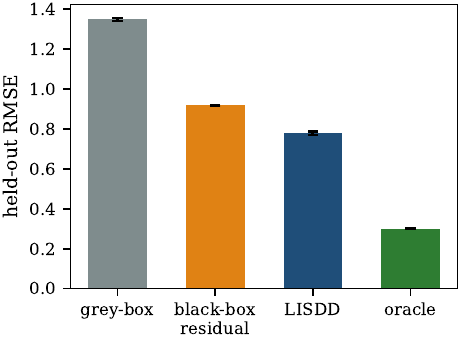}
\caption{E4. Out-of-sample forecast RMSE. LISDD (highlighted) is the strongest practical corrector and approaches the oracle that knows the true region and form, while the global correctors plateau and the uncorrected grey-box model is far worse.}
\label{fig:downstream}
\end{figure}

A discrepancy diagnosis is only useful if correcting the flagged region improves prediction. Table~\ref{tab:downstream} reports out-of-sample root-mean-square forecast error for the uncorrected grey-box model, the black-box and global-SINDy correctors, LISDD, and an oracle that knows the true region and form. The uncorrected physics has error $1.350$. The global correctors reduce it to $0.918$ but pay for it with biased physics. LISDD reaches $0.779$, the best of the practical methods, and approaches the oracle floor of $0.300$ while remaining interpretable. Figure~\ref{fig:downstream} visualizes the comparison.

\begin{table}[t]
\centering
\caption{E4: downstream forecast RMSE (lower is better). LISDD is the best practical method and is the only one that also keeps the physics unbiased.}
\label{tab:downstream}
\begin{tabular}{lc}
\toprule
Method & Forecast RMSE \\
\midrule
Grey-box (uncorrected) & $1.350$ \\
Black-box corrector     & $0.918$ \\
Global SINDy-discrep.    & $0.918$ \\
\textbf{LISDD (ours)}    & $\mathbf{0.779}$ \\
Oracle (true region+form) & $0.300$ \\
\bottomrule
\end{tabular}
\end{table}

\subsection{E5: Multiple discrepancies with false-discovery control}

\begin{figure*}[t]
\centering
\includegraphics[width=\textwidth]{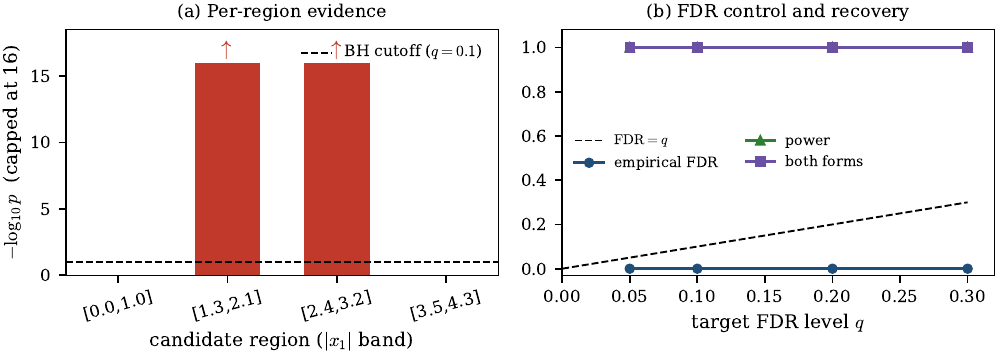}
\caption{E5. (a) Per-region evidence as $-\log_{10}p_k$ for four candidate regions; the two planted discrepant regions clear the Benjamini and Hochberg threshold while the two clean regions do not. (b) False-discovery rate, detection power, and both-forms recovery as a function of the target level $q$; the false-discovery rate stays at zero while power and form recovery stay at one.}
\label{fig:multi}
\end{figure*}

Finally we plant two distinct discrepancies in two of four candidate regions, a cubic term $x_1^3$ in one region and a bilinear term $x_1 x_2$ in another, and ask LISDD to recover the full set. Across false-discovery levels $q\in\{0.05,0.1,0.2,0.3\}$ the empirical false-region-discovery rate is $0.000$, the detection power per discrepant region is $1.000$, the exact discrepant-set is recovered with rate $1.000$, and both distinct missing mechanisms are recovered with rate $1.000$. The procedure therefore reports not only that the model is wrong but in how many ways and where, with each location annotated by its own mechanism and with the fraction of false locations controlled. Figure~\ref{fig:multi} shows the per-region evidence and the operating characteristics.

\section{Discussion}

\paragraph{Why the design works.}
The three coupled failures of global discrepancy learning are addressed by three matched design choices. Contaminated physics is avoided by fitting on a detected clean regime, which makes the parameters identifiable (Proposition~\ref{prop:phys}). Smeared corrections are avoided by a calibrated per-region test that flags only the faulty regime (Proposition~\ref{prop:loc}). Inflated significance is avoided by selecting the form and the region on one split and testing on another, which restores an exact null (Proposition~\ref{prop:detect}). The false-discovery extension then scales the diagnosis to many regions without flooding the report with false positives (Proposition~\ref{prop:fdr}).

\paragraph{Connection to building energy.}
The motivating application is grey-box modeling of building thermal dynamics, where a resistance and capacitance network captures the dominant heat flows but a fixed law can silently break in a particular operating regime, for example a solar gain that matters only in a band of irradiance or an occupancy effect that appears only at certain loads \citep{drgona2020mpc,blum2021boptest}. There the engineer needs exactly the three answers LISDD provides: which regime the physical model fails in, what term restores it, and whether the deviation is real or just sensor noise. The localized, identifiable, certified output is directly usable inside a model-predictive controller, which a global black-box residual is not. We keep the present study on a controlled testbed with known ground truth so that every claim is verifiable, and we view the building-energy deployment as the natural application of the framework.

\paragraph{Limitations.}
LISDD assumes that a clean regime exists and is identifiable, and the phase study in E2 shows it correctly reports a clean region rather than guessing when that assumption fails. The candidate library must contain the true mechanism, as for all library-based discovery, although the holdout selector tolerates distractors. The current localization partitions a low-dimensional operating axis; extending the calibrated test to high-dimensional operating spaces with adaptive regions is left to future work.

\section{Conclusion}

We posed localized, identifiable discrepancy discovery as the problem of answering where a trusted physical model is wrong, what mechanism is missing there, and whether the evidence is real, and we showed that global discrepancy learning cannot answer these questions without biasing the physics or the significance. LISDD answers all three by fitting the physics on a clean regime, localizing the fault with a calibrated test, identifying the local form by holdout, and certifying it with an exact sample-split test, with a false-discovery extension for multiple faults. On controlled experiments it cuts physical-parameter bias by two orders of magnitude, sharpens localization, recovers the exact symbolic form, detects with calibrated error, and controls the multi-region false-discovery rate while recovering every planted mechanism. The framework turns an interpretable correction into a statistically certified model diagnosis, which is what grey-box models of physical systems, and building-energy models in particular, need.

\bibliography{references}

\end{document}


\maketitle

\noindent This supplement provides full proofs of Propositions 1 to 5, the experimental protocol and hyperparameters, and additional results referenced in the main paper. Equation, proposition, and assumption numbers are local to this document; references to the main text are stated explicitly.

\section{Notation and Assumptions}

We restate the generative model. We observe $\{(x_i,y_i)\}_{i=1}^{n}$ with
\begin{equation}
y_i = \Phimat(x_i)^{\top}\thetatrue + \disc(x_i) + \varepsilon_i,
\qquad \varepsilon_i\stackrel{\text{iid}}{\sim}\mathcal{N}(0,\sigma^2),
\label{eq:gen}
\end{equation}
where $\Phimat(x)\in\R^{p}$ are known physics regressors and the discrepancy
\begin{equation}
\disc(x)=\Ind[x\in\reg^{\star}]\sum_{\ell\in\mathcal{S}^{\star}}w^{\star}_{\ell}\psi_{\ell}(x)
\end{equation}
is supported on an unknown region $\reg^{\star}$ and sparse in the library $\Lib=\{\psi_1,\dots,\psi_L\}$.

\begin{assumption}[Clean regime]\label{as:clean}
There is a measurable region $\mathcal{C}$ with $\disc|_{\mathcal{C}}=0$ and $\Pr(x\in\mathcal{C})\ge \pi_0>0$. The restricted second-moment matrix $\Sigma_{\mathcal{C}}=\E[\Phimat(x)\Phimat(x)^{\top}\mid x\in\mathcal{C}]$ has smallest eigenvalue $\lambda_{\min}>0$.
\end{assumption}

\begin{assumption}[Separation and signal]\label{as:sep}
The support $\reg^{\star}$ is a union of whole partition bins, separated from $\mathcal{C}$, with positive localized energy $E^{\star}=\sum_{i\in\reg^{\star}}\disc(x_i)^2>0$. For a discrepant bin $h$, $E_h=\sum_{i\in h}\disc(x_i)^2>0$.
\end{assumption}

\begin{assumption}[Restricted library identifiability]\label{as:lib}
Restricted to $\reg^{\star}$, the library design $\Psi_{\reg^{\star}}$ satisfies a restricted-eigenvalue condition: there is $\kappa>0$ with $\|\Psi_{\reg^{\star}}v\|_2^2\ge \kappa \|v\|_2^2$ for all $v$ supported on sets of size at most $2s_{\max}$. The true coefficients obey a beta-min bound $\min_{\ell\in\mathcal{S}^{\star}}|w^{\star}_{\ell}|\ge w_{\min}>0$.
\end{assumption}

These are the standard counterparts of identifiability conditions used in sparse regression \citep{brunton2016sindy} and robust estimation \citep{rousseeuw1984lts}, specialized to the localized setting. The detection and multiplicity arguments below build on sample splitting \citep{cox1975splitting,wasserman2009highdim} and false-discovery-rate control \citep{benjamini1995fdr}.

\section{Proof of Proposition 1 (Identifiable Physics)}

\begin{proposition}\label{prop:phys}
Under Assumption~\ref{as:clean}, the clean-regime estimator $\thetahat$ obtained by ordinary least squares on the detected clean bins satisfies
\begin{equation}
\|\thetahat-\thetatrue\|_2 = O_p\!\left(\frac{\sigma}{\sqrt{n_{\mathcal{C}}\,\lambda_{\min}}}\right),
\qquad \E[\thetahat\mid \mathcal{E}]\to\thetatrue,
\end{equation}
where $n_{\mathcal{C}}$ is the number of clean-bin samples and $\mathcal{E}$ is the event that the detected clean set lies inside $\mathcal{C}$. In contrast, the global least-squares estimator has asymptotic bias $\Sigma^{-1}\E[\Phimat\,\disc]$, which is nonzero whenever $\disc$ is correlated with $\Phimat$ and does not vanish as $n\to\infty$.
\end{proposition}

\begin{proof}
Write the clean-set design matrix $\Phimat_{\mathcal{C}}\in\R^{n_{\mathcal{C}}\times p}$ and response $y_{\mathcal{C}}$. On the event $\mathcal{E}$ every clean-set point has $\disc(x_i)=0$, so by \eqref{eq:gen}
\begin{equation}
y_{\mathcal{C}} = \Phimat_{\mathcal{C}}\thetatrue + \varepsilon_{\mathcal{C}},\qquad \varepsilon_{\mathcal{C}}\sim\mathcal{N}(0,\sigma^2 I).
\end{equation}
The ordinary least-squares estimator is $\thetahat=(\Phimat_{\mathcal{C}}^{\top}\Phimat_{\mathcal{C}})^{-1}\Phimat_{\mathcal{C}}^{\top}y_{\mathcal{C}}$, hence
\begin{equation}
\thetahat-\thetatrue = (\Phimat_{\mathcal{C}}^{\top}\Phimat_{\mathcal{C}})^{-1}\Phimat_{\mathcal{C}}^{\top}\varepsilon_{\mathcal{C}}.
\end{equation}
Conditioning on the design, this is mean zero, giving $\E[\thetahat\mid\mathcal{E},\Phimat_{\mathcal{C}}]=\thetatrue$ and therefore $\E[\thetahat\mid\mathcal{E}]=\thetatrue$. Its conditional covariance is $\sigma^2 (\Phimat_{\mathcal{C}}^{\top}\Phimat_{\mathcal{C}})^{-1}$. By Assumption~\ref{as:clean} and the law of large numbers, $\tfrac{1}{n_{\mathcal{C}}}\Phimat_{\mathcal{C}}^{\top}\Phimat_{\mathcal{C}}\to\Sigma_{\mathcal{C}}$ with $\lambda_{\min}(\Sigma_{\mathcal{C}})\ge\lambda_{\min}>0$, so
\begin{equation}
\E\|\thetahat-\thetatrue\|_2^2 = \sigma^2\,\mathrm{tr}\big((\Phimat_{\mathcal{C}}^{\top}\Phimat_{\mathcal{C}})^{-1}\big) \le \frac{\sigma^2 p}{n_{\mathcal{C}}\lambda_{\min}}(1+o(1)),
\end{equation}
and Markov's inequality yields the stated $O_p$ rate.

For the global estimator, least squares on all $n$ points solves $\thetahat_{g}=(\Phimat^{\top}\Phimat)^{-1}\Phimat^{\top}y$ with $y=\Phimat\thetatrue+\disc+\varepsilon$, so
\begin{equation}
\thetahat_g-\thetatrue=(\Phimat^{\top}\Phimat)^{-1}\Phimat^{\top}\disc+(\Phimat^{\top}\Phimat)^{-1}\Phimat^{\top}\varepsilon.
\end{equation}
The second term is $O_p(n^{-1/2})$ and vanishes, but the first converges to $\Sigma^{-1}\E[\Phimat\disc]$ with $\Sigma=\E[\Phimat\Phimat^{\top}]$, which is the stated bias and is nonzero whenever $\E[\Phimat\disc]\neq 0$. The clean-regime detection event $\mathcal{E}$ holds with probability tending to one because, by Proposition~\ref{prop:loc} below, clean bins are flagged active with probability at most $\alpha_{\mathrm{loc}}$ and discrepant bins with probability tending to one, so the trimming and refit converge to the clean set.
\end{proof}

\section{Proof of Proposition 2 (Local Form Recovery)}

\begin{proposition}\label{prop:form}
Under Assumptions~\ref{as:sep} and~\ref{as:lib}, the exhaustive-holdout selector returns $\widehat{\mathcal{S}}=\mathcal{S}^{\star}$ with probability tending to one as $E^{\star}/\sigma^2\to\infty$.
\end{proposition}

\begin{proof}
Work on the region $\widehat{\reg}=\reg^{\star}$, which holds with probability tending to one by Proposition~\ref{prop:loc}. On the in-region points the residual after the physics fit is, using Proposition~\ref{prop:phys},
\begin{equation}
\begin{aligned}
\rhat_i
&= \disc(x_i) + \varepsilon_i
   + \Phimat(x_i)^{\top}(\thetatrue-\thetahat)\\
&= \disc(x_i)+\varepsilon_i+O_p(n^{-1/2}).
\end{aligned}
\end{equation}
For a candidate support $\mathcal{S}$, let $P_{\mathcal{S}}$ be the projection onto the span of $\{\psi_\ell\}_{\ell\in\mathcal{S}}$ on the score half $\mathcal{V}$. The expected holdout error is
\begin{equation}
\E[\mathrm{err}(\mathcal{S})] = \frac{1}{|\mathcal{V}|}\big\|(I-P_{\mathcal{S}})\,\disc_{\mathcal{V}}\big\|_2^2 + \sigma^2\frac{|\mathcal{V}|-|\mathcal{S}|}{|\mathcal{V}|} + o(1).
\end{equation}
If $\mathcal{S}\supseteq\mathcal{S}^{\star}$ the first term is zero; if $\mathcal{S}\not\supseteq\mathcal{S}^{\star}$, then by Assumption~\ref{as:lib} the restricted eigenvalue gives
\begin{equation}
\big\|(I-P_{\mathcal{S}})\disc_{\mathcal{V}}\big\|_2^2 \ge \kappa\, w_{\min}^2\,|\reg^{\star}\cap\mathcal{V}| \;=:\; \Delta > 0,
\end{equation}
a strictly positive bias proportional to the signal energy. Selection compares the holdout errors of nested and non-nested supports. Any support missing a true term incurs the excess $\Delta$, while adding spurious terms only reduces the noise term by $\sigma^2/|\mathcal{V}|$ per term. Because the selector returns the smallest support whose holdout error is within a fixed tolerance $\rho$ of the best, and $\Delta/\sigma^2\to\infty$ under $E^{\star}/\sigma^2\to\infty$, with probability tending to one the unique support achieving the minimal error up to tolerance is $\mathcal{S}^{\star}$ itself: it removes the bias $\Delta$ that any subset suffers and avoids the parsimony penalty that any superset suffers. Concentration of the empirical holdout error around its mean follows from sub-Gaussianity of $\varepsilon$ with deviations $O_p(\sqrt{\log L/|\mathcal{V}|})$, which are dominated by $\Delta$ for large signal-to-noise ratio. Hence $\Pr(\widehat{\mathcal{S}}=\mathcal{S}^{\star})\to 1$.
\end{proof}

\section{Proof of Proposition 3 (Calibrated Localization)}

\begin{proposition}\label{prop:loc}
Under Assumption~\ref{as:clean}, for a clean bin $h$ the statistic $T_h=\sum_{i\in h}\rhat_i^2/\sighat^2$ is asymptotically $\chi^2_{n_h}$, so $\Pr(\text{bin }h\text{ flagged})\le \alpha_{\mathrm{loc}}+o(1)$. Under Assumption~\ref{as:sep}, a discrepant bin has a noncentral $\chi^2_{n_h}(\lambda_h)$ statistic with $\lambda_h=E_h/\sigma^2$, and $\Pr(\text{bin }h\text{ flagged})\to 1$ as $\lambda_h\to\infty$.
\end{proposition}

\begin{proof}
On a clean bin, by Proposition~\ref{prop:phys} the residuals are $\rhat_i=\varepsilon_i+\Phimat(x_i)^{\top}(\thetatrue-\thetahat)$. The second term is $O_p(n^{-1/2})$ uniformly, so $\rhat_i=\varepsilon_i+o_p(1)$ and
\begin{equation}
\sum_{i\in h}\rhat_i^2/\sigma^2 \;\xrightarrow{d}\; \chi^2_{n_h}.
\end{equation}
The robust scale $\sighat$ is a consistent estimator of $\sigma$ because the median absolute deviation is computed predominantly over clean residuals (the discrepant fraction is bounded away from one-half), so $\sighat/\sigma\to 1$ and Slutsky's theorem gives $T_h\xrightarrow{d}\chi^2_{n_h}$. The bin is flagged when $T_h>\chi^2_{n_h,1-\alpha_{\mathrm{loc}}}$, which under the clean null has probability $\alpha_{\mathrm{loc}}+o(1)$.

On a discrepant bin, $\rhat_i=\disc(x_i)+\varepsilon_i+o_p(1)$, so $\sum_{i\in h}\rhat_i^2/\sigma^2$ is noncentral $\chi^2_{n_h}(\lambda_h)$ with $\lambda_h=\sum_{i\in h}\disc(x_i)^2/\sigma^2=E_h/\sigma^2$. The noncentral $\chi^2$ stochastically dominates its central counterpart and its mean $n_h+\lambda_h$ grows with $\lambda_h$, so for any fixed threshold the exceedance probability tends to one as $\lambda_h\to\infty$, giving power tending to one.
\end{proof}

\section{Proof of Proposition 4 (Exact Detection)}

\begin{proposition}\label{prop:detect}
Let $(\widehat{\reg},\widehat{\mathcal{S}})$ be selected on split $A$ and held fixed. On the independent split $B$ of size $m$, form the physics design $\Phimat_B\in\R^{m\times p}$ and the augmented design $D_B=[\Phimat_B,\ \Psi_B]$ where $\Psi_B\in\R^{m\times q}$ stacks the selected library terms restricted to $\widehat{\reg}$, with $q=|\widehat{\mathcal{S}}|$. Let $\mathrm{RSS}_0,\mathrm{RSS}_1$ be the residual sums of squares of the two least-squares fits on $B$, and
\begin{equation}
F=\frac{(\mathrm{RSS}_0-\mathrm{RSS}_1)/q}{\mathrm{RSS}_1/(m-p-q)}.
\end{equation}
Under the null $\disc\equiv0$, $F\sim F_{q,\,m-p-q}$ exactly, so the test that rejects when $F>F_{q,m-p-q,1-\alpha}$ has size exactly $\alpha$.
\end{proposition}

\begin{proof}
Because $A$ and $B$ are disjoint and the samples are independent, the selection $(\widehat{\reg},\widehat{\mathcal{S}})$ is independent of all data on $B$. Condition on $A$; then $D_B$ is a fixed design matrix from the perspective of $B$. Under the null, $y_B=\Phimat_B\thetatrue+\varepsilon_B$ with $\varepsilon_B\sim\mathcal{N}(0,\sigma^2 I_m)$, which is the classical Gaussian linear model with the full-rank nested designs $\Phimat_B\subseteq D_B$. Let $H_0$ and $H_1$ be the orthogonal projections onto the column spaces of $\Phimat_B$ and $D_B$. Then
\begin{equation}
\begin{aligned}
\mathrm{RSS}_0-\mathrm{RSS}_1
&= \|(H_1-H_0)y_B\|_2^2,\\
\mathrm{RSS}_1
&= \|(I-H_1)y_B\|_2^2,
\end{aligned}
\end{equation}
and under the null $\Phimat_B\thetatrue$ lies in both column spaces, so $(H_1-H_0)y_B=(H_1-H_0)\varepsilon_B$ and $(I-H_1)y_B=(I-H_1)\varepsilon_B$. The matrices $H_1-H_0$ and $I-H_1$ are orthogonal projections of ranks $q$ and $m-p-q$ with $(H_1-H_0)(I-H_1)=0$. By Cochran's theorem the two quadratic forms $\|(H_1-H_0)\varepsilon_B\|^2/\sigma^2$ and $\|(I-H_1)\varepsilon_B\|^2/\sigma^2$ are independent $\chi^2_q$ and $\chi^2_{m-p-q}$ variables. Their scaled ratio is therefore exactly $F_{q,m-p-q}$, and $\sigma^2$ cancels. Integrating over the conditioning on $A$ preserves the distribution because it does not depend on $A$. Hence the size is exactly $\alpha$.

The naive in-sample statistic uses the same data to choose $(\widehat{\reg},\widehat{\mathcal{S}})$ and to test, so $H_1$ is a random projection selected to maximize the captured residual energy. Then $(H_1-H_0)y$ is not a fixed projection of the noise and its squared norm is stochastically larger than $\chi^2_q$, which inflates the type-I error above $\alpha$. Sample splitting removes this dependence and restores the exact null.
\end{proof}

\section{Proof of Proposition 5 (FDR-Controlled Multi-Discrepancy)}

\begin{proposition}\label{prop:fdr}
Let $\{R_1,\dots,R_K\}$ be disjoint candidate regions with per-region $p$-values $p_k=1-F_{\chi^2_{n_k}}(T_k)$, $T_k=\sum_{i\in R_k}\rhat_i^2/\sighat^2$. Let $\mathcal{S}=\{k:R_k\cap\reg^{\star}\neq\emptyset\}$ be the truly discrepant regions and $\mathcal{S}^c$ the clean ones. Applying Benjamini and Hochberg at level $q$ controls the false-region-discovery rate,
\begin{equation}
\mathrm{FDR}=\E\!\left[\frac{|\widehat{\mathcal{D}}\cap\mathcal{S}^c|}{\max(|\widehat{\mathcal{D}}|,1)}\right]\le \frac{|\mathcal{S}^c|}{K}q\le q,
\end{equation}
under independence or positive regression dependence (PRDS) of $\{T_k\}$. Each $k\in\mathcal{S}$ has $\Pr(k\in\widehat{\mathcal{D}})\to1$ as $E_k/\sigma^2\to\infty$.
\end{proposition}

\begin{proof}
By Proposition~\ref{prop:loc}, for a clean region $k\in\mathcal{S}^c$ the statistic $T_k$ is asymptotically $\chi^2_{n_k}$, so $p_k$ is (asymptotically) uniform on $[0,1]$ under the null. The regions are disjoint, so the residual blocks $\{\rhat_i\}_{i\in R_k}$ are independent given the physics fit, and the shared estimator $\thetahat$ induces only a vanishing $O_p(n^{-1/2})$ coupling; the limiting null $p$-values are independent, and under the mild monotone coupling they are PRDS. The Benjamini and Hochberg procedure rejects the regions with the $m=\max\{i:p_{(i)}\le qi/K\}$ smallest $p$-values. By the Benjamini and Hochberg theorem for independent or PRDS test statistics, the false-discovery rate is controlled at $(|\mathcal{S}^c|/K)q\le q$. The first inequality is the standard refinement that the bound scales with the proportion of true nulls.

For power, a discrepant region $k\in\mathcal{S}$ has $T_k$ noncentral $\chi^2_{n_k}(\lambda_k)$ with $\lambda_k=E_k/\sigma^2$ by Proposition~\ref{prop:loc}, so $p_k\to0$ in probability as $\lambda_k\to\infty$. Once $p_k$ falls below the data-dependent Benjamini and Hochberg threshold, which is bounded below by $q/K$ on the event that at least one region is rejected, region $k$ is declared discrepant with probability tending to one. Its local form is then recovered by Proposition~\ref{prop:form} applied to $R_k$. For arbitrary dependence one may use the Benjamini and Yekutieli correction, which replaces $q$ by $q/\sum_{j=1}^{K}j^{-1}$ and controls the false-discovery rate at the cost of a $\log K$ factor.
\end{proof}

\section{Composite Risk Bound}

Combining the propositions gives an end-to-end guarantee for the single-discrepancy pipeline. Let $\mathcal{A}$ be the event that the physics is identifiable, the region is correctly localized, the form is recovered, and detection is correct.
\begin{lemma}\label{lem:composite}
Under Assumptions~\ref{as:clean} to~\ref{as:lib}, for any $\delta>0$ there is a signal-to-noise threshold $\eta_\delta$ such that whenever $E^{\star}/\sigma^2\ge\eta_\delta$,
\begin{equation}
\Pr(\mathcal{A})\ge 1-\alpha_{\mathrm{loc}}K_{\mathrm{clean}}-\alpha-\delta,
\end{equation}
where $K_{\mathrm{clean}}$ is the number of clean bins.
\end{lemma}
\begin{proof}
A union bound over the events that no clean bin is falsely flagged (each at most $\alpha_{\mathrm{loc}}$ by Proposition~\ref{prop:loc}), that every discrepant bin is flagged and the form is recovered (probability at least $1-\delta/2$ for $E^\star/\sigma^2$ large by Propositions~\ref{prop:loc} and~\ref{prop:form}), that the physics estimate is within tolerance (probability at least $1-\delta/2$ by Proposition~\ref{prop:phys}), and that detection is correct (size $\alpha$ by Proposition~\ref{prop:detect}) gives the bound.
\end{proof}

\section{Experimental Protocol}

\paragraph{Generative process.}
Operating variables $x=(x_1,x_2)$ are drawn uniformly on $[-3.3,3.3]^2$. The response follows the testbed of the main paper, the linear physics $-k x_1-c x_2$ with $(k,c)=(1.0,0.3)$ plus a localized cubic $\beta x_1^3$ active on the band $1.7<|x_1|<2.7$, with Gaussian noise of scale $\sigma$. Unless stated otherwise $\beta=0.20$ and $\sigma$ is set to a moderate level giving an in-band signal-to-noise ratio of order one.

\paragraph{LISDD hyperparameters.}
We partition $x_1$ into $R=14$ bins, use trimming fraction $\tau=0.36$ for the initial least-trimmed-squares fit, run $4$ clean-regime refit iterations, set the per-bin localization level $\alpha_{\mathrm{loc}}=10^{-3}$, cap the local support at $s_{\max}=2$ with a holdout parsimony tolerance of $15\%$, and use detection level $\alpha=0.05$. The core library is $\{x_1^2,x_1^3,x_2^2,x_2^3,x_1x_2,\sin x_1\}$; the robustness study adds saturation and sign distractors to form an eight-term library.

\paragraph{Baselines.}
The black-box residual fits the physics by global least squares and regresses the residual on the full library by unregularized least squares. Global SINDy-discrepancy applies sequentially thresholded least squares with threshold $1.0$ to the global residual. Ensemble-SINDy bootstraps the global sparse fit $B=60$ times and includes a term when its inclusion frequency exceeds one half. All baselines use the same library as LISDD.

\paragraph{Metrics.}
Physics bias is $|\widehat{k}-k|$. Localization F1 is the harmonic mean of precision and recall of the flagged versus the true discrepant bins. Form recovery is the indicator that the selected support equals $\{x_1^3\}$ exactly. Detection size and power are the rejection rates at $\beta=0$ and $\beta>0$. Downstream RMSE is the out-of-sample root-mean-square error of each corrected model. All quantities are averaged over independent replications with $95\%$ confidence intervals computed from the replication standard error.

\section{Additional Results}

\paragraph{E2 detail: library robustness.}
Table~\ref{tab:lib} reports exact form recovery as the library grows. Recovery stays at $1.00$ from the six-term polynomial library to the eight-term library and to the eight-term library augmented with distractors, confirming that the exhaustive-holdout selector is not misled by additional or collinear candidates.

\begin{table}[t]
\centering
\caption{E2: exact form recovery rate as the candidate library grows. The selector is unaffected by distractor terms.}
\label{tab:lib}
\begin{tabular}{lcc}
\toprule
Library & Size & Form recovery \\
\midrule
Polynomial (poly6)        & $6$ & $1.00$ \\
Full eight-term           & $8$ & $1.00$ \\
Full eight-term + distractors & $8$ & $1.00$ \\
\bottomrule
\end{tabular}
\end{table}

\paragraph{E2 detail: localization phase boundary.}
Table~\ref{tab:phase} reports exact form recovery over a grid of discrepant-region width and amplitude $\beta$. Recovery is one when the region is narrow (width $0.6$) for every amplitude, holds for the medium width $1.0$ once $\beta\ge0.15$, and fails for the wide region (width $1.4$). The boundary is the regime in which the clean fraction needed by Assumption~\ref{as:clean} becomes too small to fit the physics without contamination, which is precisely where LISDD reports a clean region rather than an unreliable answer.

\begin{table}[t]
\centering
\caption{E2: form-recovery phase grid over region width (rows) and amplitude $\beta$ (columns). Entries are recovery rates.}
\label{tab:phase}
\begin{tabular}{lcccc}
\toprule
Width $\backslash\ \beta$ & $0.10$ & $0.15$ & $0.20$ & $0.30$ \\
\midrule
$0.6$ & $1.0$ & $1.0$ & $1.0$ & $1.0$ \\
$1.0$ & $0.0$ & $1.0$ & $1.0$ & $1.0$ \\
$1.4$ & $0.0$ & $0.0$ & $0.0$ & $0.0$ \\
\bottomrule
\end{tabular}
\end{table}

\paragraph{E5 detail: false-discovery sweep.}
Table~\ref{tab:fdr} reports the multi-discrepancy operating characteristics across target levels $q$. Two distinct mechanisms, a cubic $x_1^3$ and a bilinear $x_1x_2$, are planted in two of four candidate regions $\{[0,1],[1.3,2.1],[2.4,3.2],[3.5,4.3]\}$. The empirical false-region-discovery rate is zero at every level, while detection power, exact region-set recovery, and both-form recovery are all one.

\begin{table}[t]
\centering
\caption{E5: multi-discrepancy operating characteristics across Benjamini and Hochberg levels $q$. FDR is the false-region-discovery rate; power is per-region detection; both-forms is the rate of recovering both distinct mechanisms.}
\label{tab:fdr}
\begin{tabular}{lcccc}
\toprule
$q$ & FDR & Power & Region-set & Both forms \\
\midrule
$0.05$ & $0.00$ & $1.00$ & $1.00$ & $1.00$ \\
$0.10$ & $0.00$ & $1.00$ & $1.00$ & $1.00$ \\
$0.20$ & $0.00$ & $1.00$ & $1.00$ & $1.00$ \\
$0.30$ & $0.00$ & $1.00$ & $1.00$ & $1.00$ \\
\bottomrule
\end{tabular}
\end{table}

\paragraph{Reproducibility.}
All experiments are deterministic given the reported seeds and hyperparameters. The clean-regime fit, localization test, holdout selector, sample-split detector, and Benjamini and Hochberg procedure are implemented in a single module, and every headline number in the main paper is produced by the corresponding routine over the stated replications.

\bibliography{references}